\documentclass[journal=apchd5,manuscript=article]{achemso}

\usepackage{chemformula} 
\usepackage[T1]{fontenc} 
\usepackage{graphicx}
\usepackage[normalem]{ulem}	

\author{Luca Sortino}
\affiliation{Department of Physics and Astronomy, University of Sheffield, Sheffield, S3 7RH}
\email{l.sortino@sheffield.ac.uk}
\author{Matthew Brooks}
\affiliation{Department of Physics, University of Konstanz, D-78464 Konstanz, Germany}
\author{Panaiot G. Zotev}
\author{Armando Genco}
\affiliation{Department of Physics and Astronomy, University of Sheffield, Sheffield, S3 7RH}
\author{Javier Cambiasso}
\author{Sandro Mignuzzi}
\affiliation{The Blackett Laboratory, Department of Physics, Imperial College London, London, SW7 2BW}
\author{Stefan A. Maier}
\affiliation{The Blackett Laboratory, Department of Physics, Imperial College London, London, SW7 2BW}
\alsoaffiliation{Chair in Hybrid Nanosystems, Nanoinstitute M{\"u}nich, Faculty of Physics, Ludwig-Maximilians-Universit{\"a}t M{\"u}nchen, 80539 M{\"u}nich, Germany}
\author{Guido Burkard}
\affiliation{Department of Physics, University of Konstanz, D-78464 Konstanz, Germany}
\author{Riccardo Sapienza}
\affiliation{The Blackett Laboratory, Department of Physics, Imperial College London, London, SW7 2BW}
\author{Alexander I. Tartakovskii}
\affiliation{Department of Physics and Astronomy, University of Sheffield, Sheffield, S3 7RH}
\email{a.tartakovskii@sheffield.ac.uk}

\title{Dielectric nano-antennas for strain engineering in atomically thin two-dimensional semiconductors}


\begin{document}
	
	\begin{tocentry}
		\includegraphics[width=0.7\linewidth]{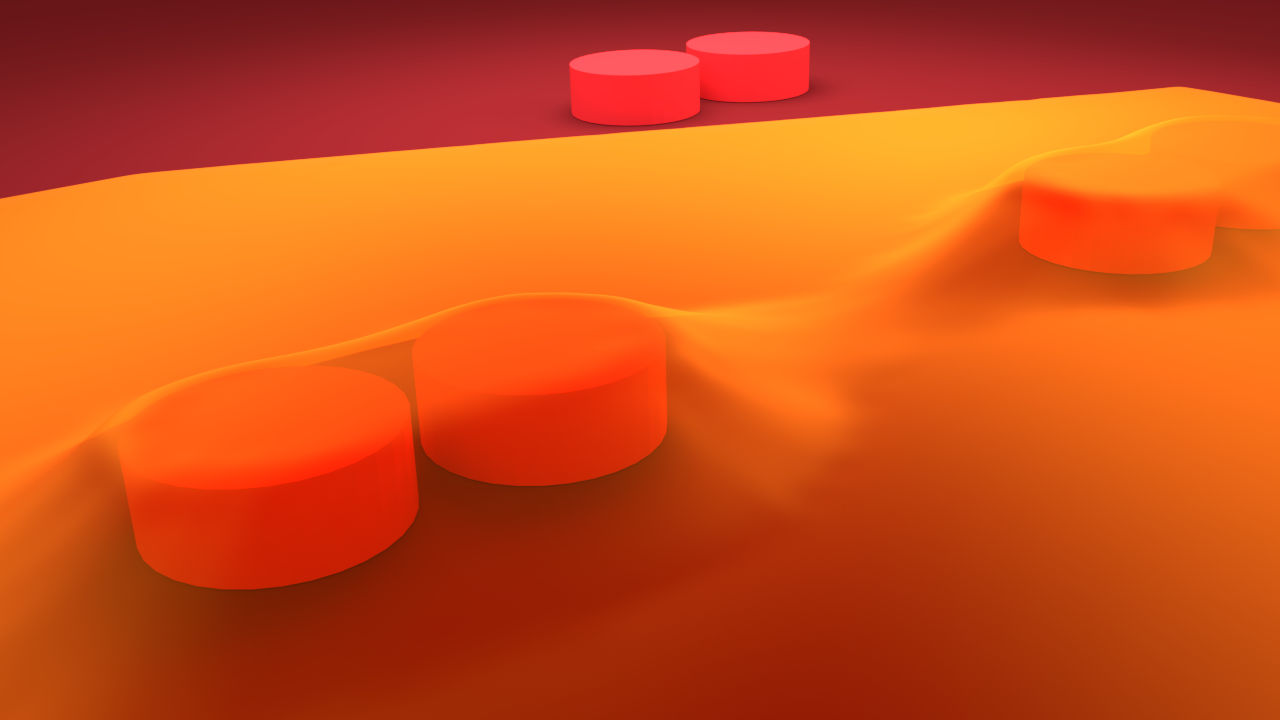}
	\end{tocentry}
	
	\pagebreak
	\begin{abstract}
		Atomically thin two-dimensional semiconducting transition metal dichalcogenides (TMDs) can withstand large levels of strain before their irreversible damage occurs. This unique property offers a promising route for control of the optical and electronic properties of TMDs, for instance by depositing them on nano-structured surfaces, where position-dependent strain can be produced on the nano-scale. Here, we demonstrate strain-induced modifications of the optical properties of mono- and bilayer TMD WSe$_2 $ placed on photonic nano-antennas made from gallium phosphide (GaP). Photoluminescence (PL) from the strained areas of the TMD layer is enhanced owing to the efficient coupling with the confined optical mode of the nano-antenna. Thus, by following the shift of the PL peak, we deduce the changes in the strain in WSe$_2$ deposited on the nano-antennas of different radii. In agreement with the presented theory, strain up to $\approx 1.4 \%$ is observed for WSe$_2$ monolayers. We also estimate that $>3\%$ strain is achieved in bilayers, accompanied with the emergence of a direct bandgap in this normally indirect-bandgap semiconductor. At cryogenic temperatures, we find evidence of the exciton confinement in the most strained nano-scale parts of the WSe$_2$ layers, as also predicted by our theoretical model. Our results, of direct relevance for both dielectric and plasmonic nano-antennas, show that strain in atomically thin semiconductors can be used as an additional parameter for engineering light-matter interaction in nano-photonic devices.
	\end{abstract}
	

	\pagebreak
	
	Control over lattice distortions in semiconductors, or strain-engineering, offers valuable tools for tailoring their electronic and optical properties. 
	The exceptional flexibility of atomically thin two-dimensional (2D) crystals \cite{Lee2008} has opened the way to employ mechanical deformation for engineering their physical properties \cite{Guinea2010,Duerloo2012,Wu2014c,Roldan2015,Amorim2015,Dai2019}, and has made them promising for applications in flexible electronics \cite{Akinwande2014,Mueller2018}.
	Among 2D semiconductors, the family of transition metal dichalcogenides \cite{Manzeli2017} has shown attractive properties including an indirect-to-direct bandgap transition in monolayers \cite{Mak2010,Splendiani2010}, tightly bound excitons \cite{Wang2017} and coupling between the spin and valley degrees of freedom \cite{Xu2014}.
	Single layers of TMDs, only three atoms thick, have been observed to withstand large strain levels ($ >$10\%)\cite{Bertolazzi2011} before fracture,
	offering a unique possibility to engineer their electronic band structure using strain \cite{Yun2012,Johari2012,Chang2013a,Guzman2014,Kormanyos2015a}. This approach has been successfully applied for modifying their optical properties \cite{Conley2013,Castellanos-Gomez2013,Feierabend2017,Aslan2018a,Schmidt2016b,Tedeschi2019}, tuning carrier mobilities \cite{Liu2019} and controlling the charge transport \cite{Manzeli2015,DeSanctis2018}. 
	Furthermore, spatial variation of the strain in TMD layers, occurring due to imperfections of fabrication methods or introduced intentionally, translates into a position dependent bandgap \cite{Shin2016,Blundo2020}.
	This was shown to result in periodically modulated optical properties in monolayer MoS$_2$ placed on pre-patterned substrates \cite{Li2015}, as well as to give rise to funnelling of excitons into the areas with a larger tensile strain \cite{Feng2012,San-Jose2016}, where the lowest exciton energy is achieved. Finally, in  mono- and bi-layer WSe$_2$ and WS$_2$ placed on silica or polymer nano-pillars, single photon emitting defects were readily observed, whose origin was thus linked with the strain induced in the few-layer semiconductor \cite{Branny2016,Palacios-Berraquero2016c}.

	While initial demonstrations of strained 2D semiconductors in electronic devices have been reported \cite{Liu2019,Gant2019}, the use of strain to control the optical properties of TMDs in nano-photonic systems has been largely unexplored \cite{Shiue2017}.
	One of the approaches to engineer light-matter interaction relies on the use of nano-antennas \cite{Schuller2010,Novotny2011,Koenderink2017}, which can confine light in volumes below the diffraction limit \cite{Baranov2018}.
	The strong confinement of the electric field at the hot-spots of the nano-antenna, enhances the light-matter interaction for the emitters positioned within the volume of its optical mode \cite{Sortino2019,Akselrod2015,Wang2016c}.
	While these phenomena have been extensively studied for surface plasmons in metals \cite{Maier,Giannini2011}, recently high-refractive-index dielectric nano-photonic structures \cite{Kuznetsov2016} have rapidly gained attention as they offer low losses \cite{Cambiasso2018,Caldarola2015} and both electric and magnetic Mie-type optical resonances \cite{Kuznetsov2012}. Recent reports of dielectric nano-antennas interfaced with 2D TMDs showed modified PL directionality in monolayer MoS$ _2 $  coupled to silicon nano-rods \cite{Cihan2018}, as well as  strong PL and Raman signal enhancement for WSe$_2$ placed on top of GaP dimer nano-pillars \cite{Sortino2019}. In this work, we investigate the role of strain in the latter case of the dimer structures. Compared with a more trivial case of single pillars, which can also be used to provide local strain \cite{Palacios-Berraquero2016c,Branny2016}, the dimer nano-pillars provide stronger light-matter interaction with more interesting photonic properties \cite{Sortino2019}, thus motivating our study.
	
	\begin{figure*}[t!]
		\centering
		\includegraphics[width=1\linewidth]{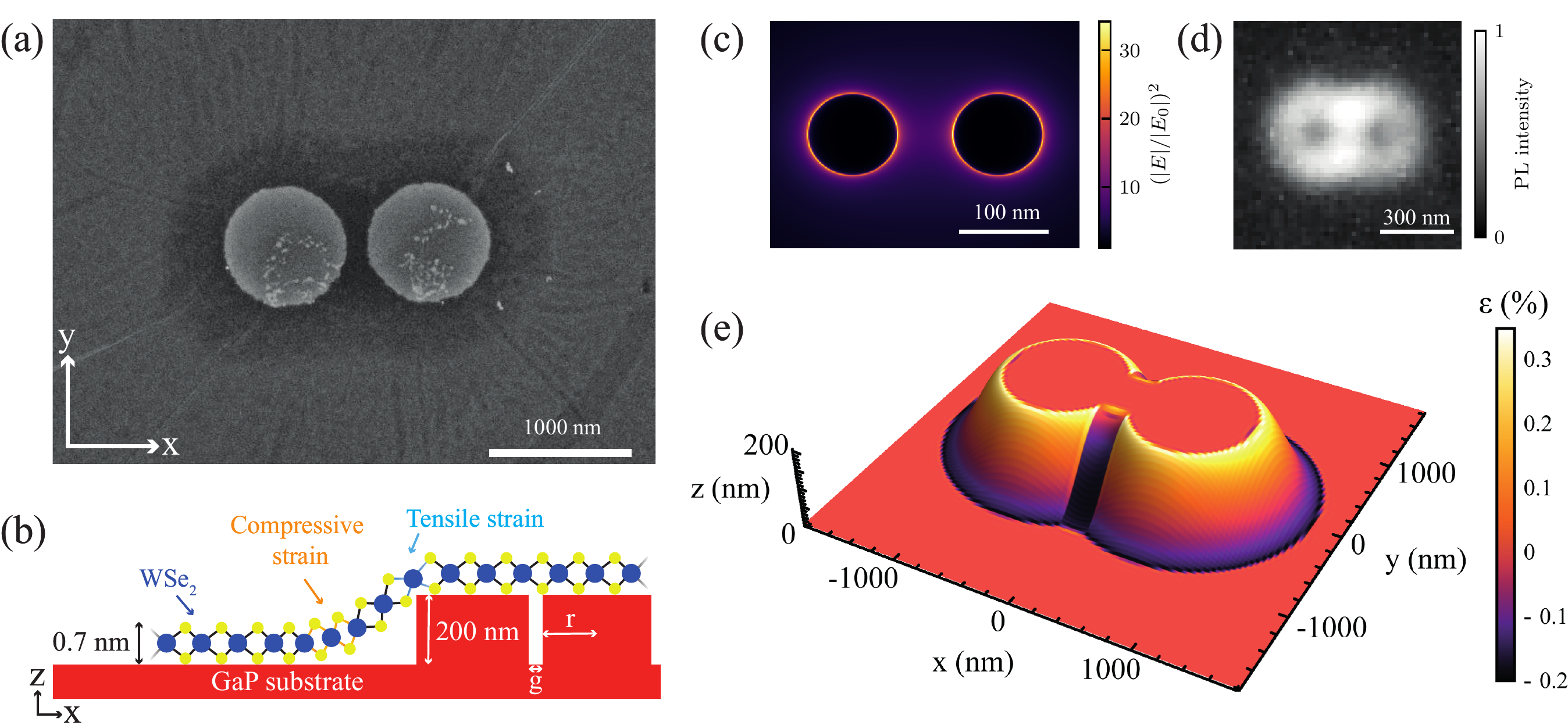}
		\linespread{1}
		\caption{
			(a) Scanning electron microscope image of a GaP dimer nano-antenna (top view) with a radius $r$=400 nm, covered with 2L-WSe$_{2} $. Scale bar: 1 $ \mu $m. The darker area around the nano-pillars is the suspended 2D layer stretching from the substrate up to the top surface of the nano-antenna.
			(b) Schematic of a GaP dimer nano-antenna shown in red with a monolayer of WSe$_2 $ transferred on its top. The areas of high compressive and tensile strain are labelled. 
			(c) The ratio of the intensities of the scattered and incident electro-magnetic waves shown as the ratio of the corresponding electric fields $(|E|/|E_0|)^2$ for a GaP dimer nano-antenna with $r$=50 nm under 685 nm unpolarized excitation. The presented field distribution is calculated at the top of the dimer ($z$=200 nm) with a finite-difference time-domain software (Lumerical Inc.). Scale bar: 100 nm. 
			(d) Photoluminescence image of a GaP nano-antenna ($ r=300 $ nm) covered with monolayer WSe$_2$. The brightest PL is observed where the tensile strain and near-field enhancement are co-located and maximized. Scale bar: 600 nm.
			(e) Calculated strain $ \varepsilon $ for a monolayer WSe$_{2}$ placed on top of a dimer nano-antenna with $ r=500 $ nm. The middle of the gap corresponds to $x$=0 and $y$=0.}
		\label{fig:fig1}
	\end{figure*}

	Contrary to previous approaches employing bending apparatuses \cite{Conley2013}, piezoelectric substrates \cite{Hui2013} or relying on the difference of the thermal expansion coefficients \cite{Plechinger2015,Ahn2017}, here we demonstrate that dimer nano-antennas can be used to strain a 2D semiconducting layer, while at the same time the PL from the strained material is enhanced via the coupling with the optical mode of the nano-antenna \cite{Sortino2019}. 
	We study both experimentally and theoretically monolayers (1L) and bilayers (2L) of TMD WSe$_2 $ transferred on closely spaced ($<$100 nm) pairs of nano-pillars, referred to as dimers \cite{Cambiasso2017a,Sortino2019} below (see Figs.\ref{fig:fig1}a-d). 
	The co-location between WSe$_2$ tensile deformation and the maximum of the near-field enhancement of the antenna at the edges of the top surface of the pillars (Fig.\ref{fig:fig1}b-e), allows clear observation of the strain-induced band structure modifications in PL. 
	We demonstrate a dependence of the amount of strain induced in the TMD layer by nano-antennas of different radii. 
	With this approach we observe the tuning of the intralayer exciton in monolayer WSe$_2$ exceeding 50 meV, corresponding to the uniaxial tensile strain up to $ \approx1.4\%$ \cite{Niehues2018a}. 
	These findings are supported by our model based on the continuum-mechanical plate theory approach \cite{Brooks2018,Landau}, which we use to calculate the strain distribution in 1L and 2L-WSe$_2$ placed on the nano-antennas. 
	Using this theory we predict higher strain in the bilayers than in monolayers placed on the nano-antennas of the same radius. 
	Tensile strain $> 3\% $ is calculated for 2L-WSe$_2$ placed on the nano-antenna with $ r=50 $ nm, for which we experimentally observe abrupt changes in the PL spectrum consistent with the transition to the direct bandgap. 
	From the calculated topography model, we deduce the position-dependent band structure induced in the WSe$_2 $ and predict occurrence of   strain-induced potential wells at the nano-pillar edges, which can trap the excitons. This behaviour is confirmed in the cryogenic PL measurements of 2L-WSe$_{2} $, which exhibits localization of the excitonic PL, consistent with the predicted potential shape. 
	Further to demonstrating tuning of strain in TMDs using all-dielectric nano-antennas of various dimensions, our findings have direct relevance to TMDs coupled to plasmonic structures, where similarly both the strain and the enhanced optical fields may be co-located \cite{Cai2018,Luo2018}.

	\section*{Results}
	
	\textbf{Strained WSe$_2 $ coupled with GaP nano-antennas}
	Fig.\ref{fig:fig1}a shows an electron microscopy image of the top view of a GaP dimer nano-antenna with an atomically thin WSe$ _2 $ bilayer deposited on top. Owing to its elasticity, the thin semiconductor crystal stretches from the substrate up to the topmost edge of the nano-antenna, without compromising its integrity. Fig.\ref{fig:fig1}b shows a schematic diagram of a WSe$_2$ monolayer (0.7 nm thick) transferred on top of a GaP dimer nano-antenna. We use nano-antennas with a height of 200 nm, a radius varying from 50 to 500 nm and a gap from 50-100 nm. Due to the height mismatch between the substrate and the top surface of the antenna, the WSe$ _2 $ layer experiences a local compressive or tensile deformation, as shown in Fig.1b,e. The optical response of a dimer nano-antenna exhibits spectrally broad resonances \cite{Cambiasso2017a,Sortino2019} (see calculated spectra in Supplementary Note I). Fig.\ref{fig:fig1}c shows the spatial intensity distribution of the scattered light of a nano-antenna illuminated with a planar wave at the top surface of a GaP dimer nano-antenna under unpolarized excitation. We present this scattered intensity as $ (|E|/|E_0|)^2 $, where $ E $ is the electric field in the scattered wave, and $ E_0 $ is the electric field of the plane wave illuminating the structure. 
	Maxima of the field intensity are located at the edges of each nano-pillar and inside the gap region between them. The transferred 2D semiconductor layer stretches over the nano-structure, fully overlapping with the confined optical modes \cite{Sortino2019}. 
	An optical microscope image of 1L-WSe$_2$ PL emission on top of a single GaP dimer nano-antenna, with $ r=300 $ nm and $ g = 65 $ nm, is shown in Fig.\ref{fig:fig1}d (see Methods and Ref.\cite{Alexeev2017} for more details). In this image it is possible to resolve the spatial distribution of the PL signal which closely resembles that of the enhanced field in Fig.\ref{fig:fig1}c.

	We describe the nano-scale distortions in the 2D layer with an approach based on the continuum-mechanical plate theory \cite{Brooks2018} by modelling the pressure experienced by the stretched semiconductor in the out-of-plane direction (see Supplementary Note II for the full description of the model).
	From the analytically calculated function for a single pillar, we interpolate to the low-symmetry dimer structure and calculate the spatial distribution of strain introduced in the stretched semiconductor crystal.
	The resulting strain topography for a 1L-WSe$_2 $ placed on top of a dimer nano-antenna is shown in Fig.\ref{fig:fig1}e. The tensile strain reaches its maximum at the topmost edge of each nano-pillar, at a height $z= 200 $ nm, co-located with the largest near-field enhancement produced by the underlying nano-antenna. The regions of compressive strain are located where the 2D layer touches the substrate, at $z\approx0$, and inside the gap region between the pillars. 
	
	\begin{figure*}[t!]
		\centering
		\includegraphics[width=1\linewidth]{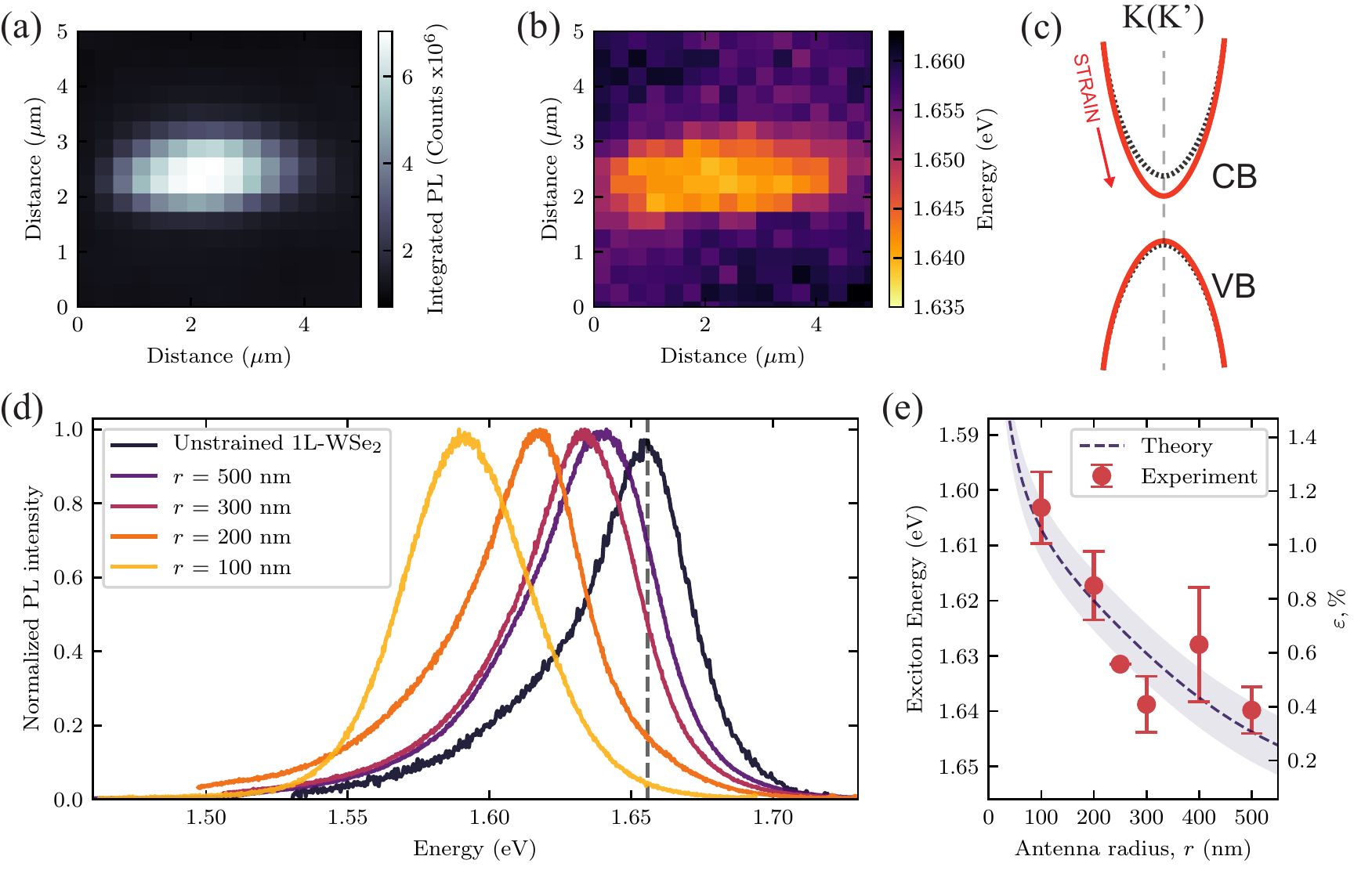}
		\linespread{1}
		\caption{(a) PL intensity map for 1L-WSe$_{2}$ placed on a GaP nano-antenna with $ r =$ 300 nm measured using a micro-PL set-up. (b) The same map as in (a), but showing the exciton peak energy distribution. (c) Schematics showing how the band-structure around the $K$ valley in 1L-WSe$_{2} $ is modified under tensile strain (red). The band-structure for the unstrained 1L-WSe$_{2} $ is shown in black. (d) Normalized PL spectra for 1L-WSe$_{2} $ placed on top of nano-antennas with different radii $r$. The red-shift observed for  decreasing $r$ corresponds to increasing strain in WSe$_2$. (e) Red circles show the average PL peak positions for 1L-WSe$ _2 $ placed on nano-antennas with different radius and corresponding uniaxial strain deduced using the experimental gauge factor from Ref.\cite{Niehues2018a}. The error bars represent the standard deviation between different dimers of the same radius. The dashed line shows the strain values calculated using our theoretical model. The shaded region corresponds to a value of $\pm$0.1\% of strain around the theoretical curve, equal to the average standard deviation found in experiments.}
		\label{fig:fig2}
	\end{figure*}
	
	In what follows, we show that the co-location of the maximum of the confined optical mode and that of the tensile deformation in the 2D layer, allows us to directly probe the strain-induced band structure renormalization by monitoring the enhanced excitonic room temperature PL emission for both 1L and 2L-WSe$_2 $. At cryogenic temperature, we show the exciton PL localization in 2L-WSe$_2 $ in the regions of the largest tensile strain where the reduction of the exciton energy is the strongest, as predicted by our theoretical model.
	
	\textbf{Strain tuning in 1L-WSe$_{2}$ coupled to GaP nano-antennas}
	We studied the emission from the same nano-antenna shown in Fig.\ref{fig:fig1} in a micro-PL setup at room temperature using PL mapping with a micron resolution (see Methods), as shown in Fig.\ref{fig:fig2}a. The sample is excited at $ \lambda_{exc} =$ 685 nm, below the absorption edge of GaP, and thus this light is only absorbed by the WSe$_2$ layer. As expected, a large enhancement in the overall WSe$_2 $ PL intensity is correlated with the location of the nano-antenna. Note, that in contrast to Fig.\ref{fig:fig1}d, here the resolution is limited by the excitation spot size ($r \approx1$ $\mu$m). 
	As shown in Fig.\ref{fig:fig2}b, we extract the peak maxima from the PL map in Fig.\ref{fig:fig2}a and observe a prominent red-shift of the intralayer exciton (also referred to as 'A exciton') peak on the nano-antenna. This effect is expected from the reduction of the WSe$_2$ bandgap under tensile strain.
	As schematically shown in Fig.\ref{fig:fig2}c, the direct optical transition in 1L-WSe$_2$ is located at the K(K') point of the Brillouin zone \cite{Splendiani2010,Mak2010}. Under the increasing tensile strain, the bandgap is expected to reduce \cite{Schmidt2016b,Aslan2018a} mainly due to the lowering of the conduction band minimum, as a consequence of the change in the electronic orbital overlap following the the changes in the atomic bond lengths and angles \cite{Chang2013a,Johari2012,Kormanyos2015a}. 
	Strain induced by underlying nano-structures has previously been measured using Raman enhancement in graphene on plasmonic nano-antennas\cite{Heeg2013,Heeg2013a}. As we show in Supplementary Note III, we collected the Raman scattering signal from the strained 1L-WSe$_2$ and observe a shift of the Raman peak, as expected for tensile strain \cite{Desai2014}.
	
	From our theoretical description we have derived a dependence of the tensile strain maximum, located at the nano-pillar edges, on the nano-antenna size (see Supplementary Note II). We found that by reducing the nano-pillar radius an increased strain is introduced in the 2D layer. We confirm this trend experimentally in Fig.\ref{fig:fig2}d which shows the normalized 1L-WSe$_2$ PL spectra for the TMD coupled to nano-antennas of different radii (see also Supplementary Note IV for raw PL spectra and more PL characterization data). For comparison we also show the PL spectrum for WSe$_2$ on a planar GaP substrate, where the intralayer exciton peak is marked by a dashed line.
	When the nano-pillar radius is reduced, the larger red-shift in the exciton PL peak confirms that the semiconducting layer is experiencing an increased strain, as predicted by our model.

	By using the previously experimentally observed gauge value for the WSe$_2$ intralayer A exciton red-shift as a function of strain of $ -49\pm 2$ meV/$\%$ \cite{Niehues2018a}, we extract the uniaxial tensile strain magnitude for WSe$_2$ placed on the nano-antennas with different radii.
	As shown in Fig.\ref{fig:fig2}e, we find tensile strain values up to $ \approx1.4\%$ as the nano-antenna radius is reduced, comparable to the estimate from our theoretical model shown with the dashed line.
	There is a variation in the peak position measured on different pillars. This effect can be attributed to the non-uniform coverage of the antenna surface by the monolayer and from the local disorder introduced in the WSe$_2 $ layer for instance from wrinkles, contamination etc. Further variation may occur due to imperfections in fabrication of the nano-antennas. These effects are also responsible for additional broadening of the PL linewidth, which is otherwise expected to reduce in WSe$ _2$ monolayers under uniform strain \cite{Niehues2018a}.

\begin{figure*}[t!]
		\includegraphics[width=0.65\linewidth]{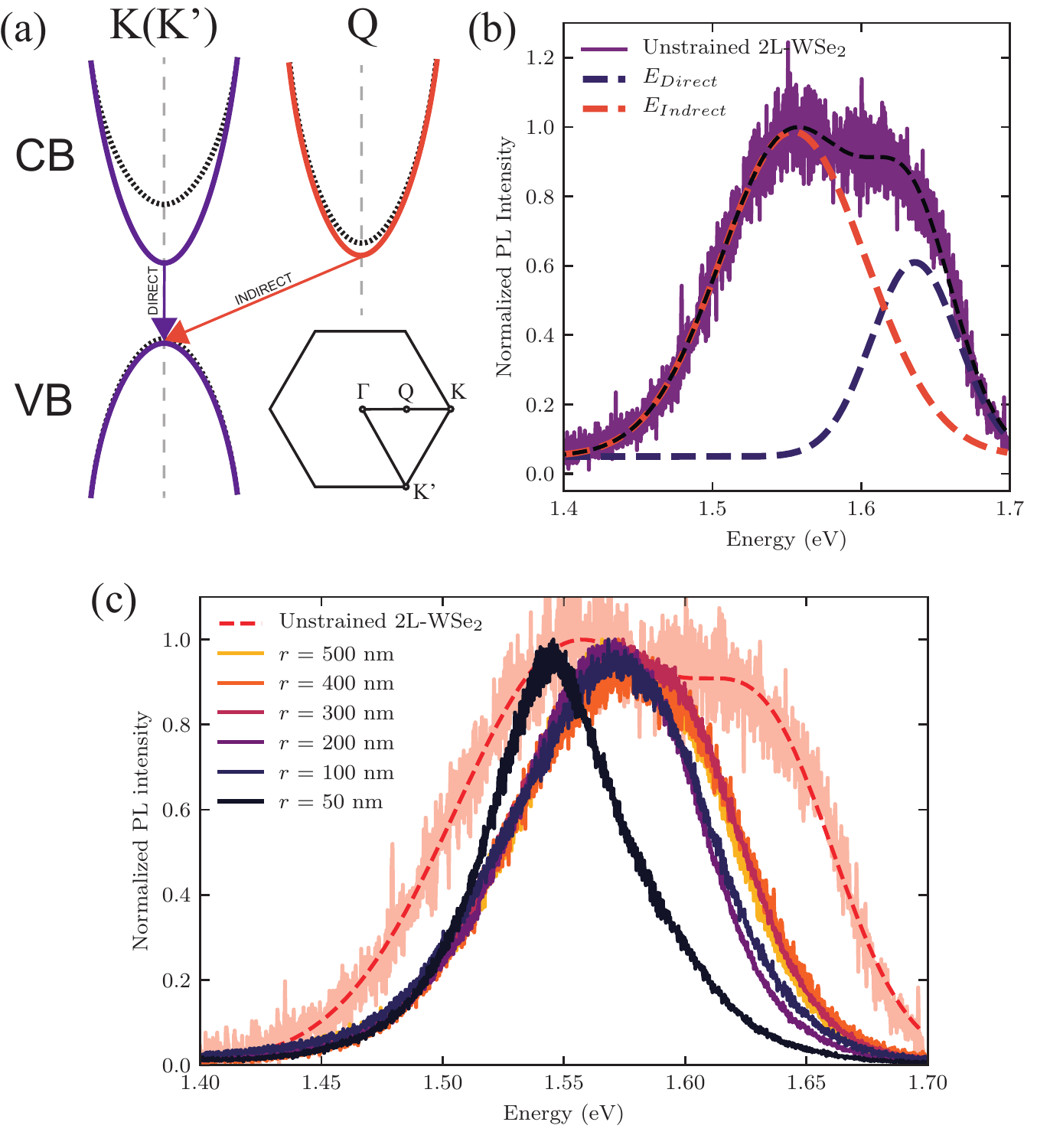}
		\linespread{1}
		\caption{(a) Schematic of the band structure modification in 2L-WSe$_{2} $ under tensile strain. The black dashed line shows the case of the unstrained 2L-WSe$_{2} $, where the lowest energy radiative recombination occurs as a result of the momentum-indirect $ Q\rightarrow K $ transition. Purple (orange) lines show the modification of the $K$($Q$) valley energies  under tensile strain. Arrows show momentum-direct (purple) and indirect (orange) transitions giving rise to PL. (b) PL spectrum of unstrained 2L-WSe$_2 $ placed on a  planar GaP substrate. Fitting reveals individual PL peaks associated with the indirect ($Q\rightarrow K$,  $E_{Indirect}$, orange) and the direct ($K\rightarrow K$,  $E_{Direct}$, purple) transitions. (c) Normalized PL spectra measured for 2L-WSe$_{2} $ on top of GaP nano-antennas with different radii and deposited on planar GaP substrate (dashed line).}
		\label{fig:fig3}
	\end{figure*}

	\textbf{Direct bandgap transition in 2L-WSe$_{2} $}
	As schematically shown in Fig.\ref{fig:fig3}a, using the band-structure picture without excitonic effects, contrary to the monolayer case, a WSe$_2 $ bilayer is an indirect bandgap semiconductor. 
	It exhibits two main recombination pathways at room temperature, a momentum direct (higher energy) and indirect (lower energy) transition, respectively involving the K and Q points of the conduction band \cite{Yun2012}. 
	For unstrained 2L-WSe$_2$ (black dashed line) the lowest energy transition is the phonon-assisted recombination between the conduction band minimum at the Q point and the valence band maximum at the K point.
	Under tensile strain (coloured lines in Fig.\ref{fig:fig3}a), similar to the monolayer case, the conduction band minimum at the K point exhibits a reduction in energy, leading to a transition from the indirect to direct bandgap in highly strained 2L-WSe$_2 $ \cite{Desai2014,Wu2018a}. In previous reports, there were contrasting interpretations for the dependence of the Q valley energy on strain \cite{Desai2014,Wu2018a}. In Fig.\ref{fig:fig3}a we use the interpretation from Ref.\cite{Wu2018a}, thus assuming that the Q valley energy would reduce under the tensile strain. However, the tuning of the Q valley is found to be negligible in the range of strain investigated in our samples (see more detailed PL spectra analysis in Supplementary Note V). The bandgap behaviour is thus determined by the stronger shift of the K valley to lower energy under tensile strain.
	Note, that in the exciton picture, the behaviour described above will translate in the corresponding energy shifts of the momentum-bright K-K exciton and the momentum-dark Q-K exciton.
	
	Fig.\ref{fig:fig3}b shows a room temperature PL spectrum for the unstrained 2L-WSe$_{2} $ placed on a flat GaP substrate. In contrast to what found for the monolayer, the bilayer exhibits a broad spectrum composed of two peaks, attributed to the momentum indirect $ E_{Indirect} = 1.55$ eV and momentum direct $ E_{Direct} = 1.63 $ eV exciton transitions \cite{Yun2012,Zhao2013}. The corresponding individual peaks obtained from the fitting of the PL spectrum are shown with dashed lines in Fig.\ref{fig:fig3}b.
	
	In analogy with the analysis carried out for WSe$_2 $ monolayers, we collected the PL emission of the strained bilayer placed on top of the nano-antennas with different radii. The normalized PL spectra measured for the nano-antennas of different radii are shown in Fig.\ref{fig:fig3}c (see also Supplementary Fig.S5).
	For the strained WSe$_2 $ bilayer, the presence of two transitions leads to a less evident strain-induced shift of the PL spectra. 
	However, by comparing the spectral shape to that of the unstrained case we observe that, as the nano-antenna radius is reduced, the higher energy side of the spectrum shows a consistent shift to lower energies, following an abrupt change of the PL spectrum from that on the flat GaP (dashed red line) to the spectrum measured on a pillar with $ r=500 $ nm (yellow).
	The shift is accompanied with a reduction of the linewidth related to the reduction of the exciton-phonon coupling in WSe$_2 $ under strain \cite{Niehues2018a}. 
	We attribute the red-shift under increasing tensile strain to the reduction in energy of the $ K\rightarrow K $ transition, as we observed for the monolayer case.
	Moreover, the PL lineshape exhibits a further abrupt change when the 2L-WSe$_2 $ is deposited on top of a GaP dimer nano-antenna with $ r=50 $ nm. 
	This spectral shape and peak position resemble those reported in Ref.\cite{Desai2014,Wu2018a} where they were attributed to a transition to the direct bandgap, occurring due to the red-shift of the high-energy peak at $ E_{Direct} $ (see also Supplementary Note V).
	Our theoretical model predicts strain values $ >3\%$ for 2L-WSe$_2 $ placed on the nano-antennas with $ r=50 $ nm (see Supplementary Fig.S2). This value is comparable with the reported threshold for the direct bandgap transition observed in WSe$_2 $ bilayers \cite{Wu2018a}.
	
	Note, that for the pillars of the same radius, a higher strain is introduced in the bilayer compared with a monolayer, due to the larger rigidity and larger mass of the former. Thus, a stronger bandgap tuning is observed in a bilayer. A similar effect has been reported in suspended 1L and 2L-MoS$ _2 $ under the strain introduced by an AFM probe \cite{Manzeli2015}.

	\textbf{Exciton confinement in the strain-induced potential}
	Controlled nano-scale deformations observed in Fig.1-3 allow the control of exciton motion in 2D semiconductors by means of band structure modification \cite{Feng2012,San-Jose2016}. 
	For WSe$_{2} $, the application of tensile strain results in the increase of the valence band edge energy and reduction of the conduction band edge energy at the K point. This effect leads to so-called exciton funnelling \cite{Brooks2018,Feng2012}, where a strain gradient directs the excitonic population towards the strain-induced potential well. 
	This effect has been proposed to play a role in the population of the single photon emitting centres in WSe$_2 $ at cryogenic temperatures \cite{Palacios-Berraquero2016c,Branny2016}.
	In order to elucidate the role of the confinement produced by the strain, we probed the spatial dependence of the PL emission of 2L-WSe$_2 $. The experiments are carried out at a cryogenic temperature of 4 K where the effect of the strain-induced confinement is most pronounced.
	
\begin{figure*}
		\centering
		\includegraphics[width=0.77\linewidth]{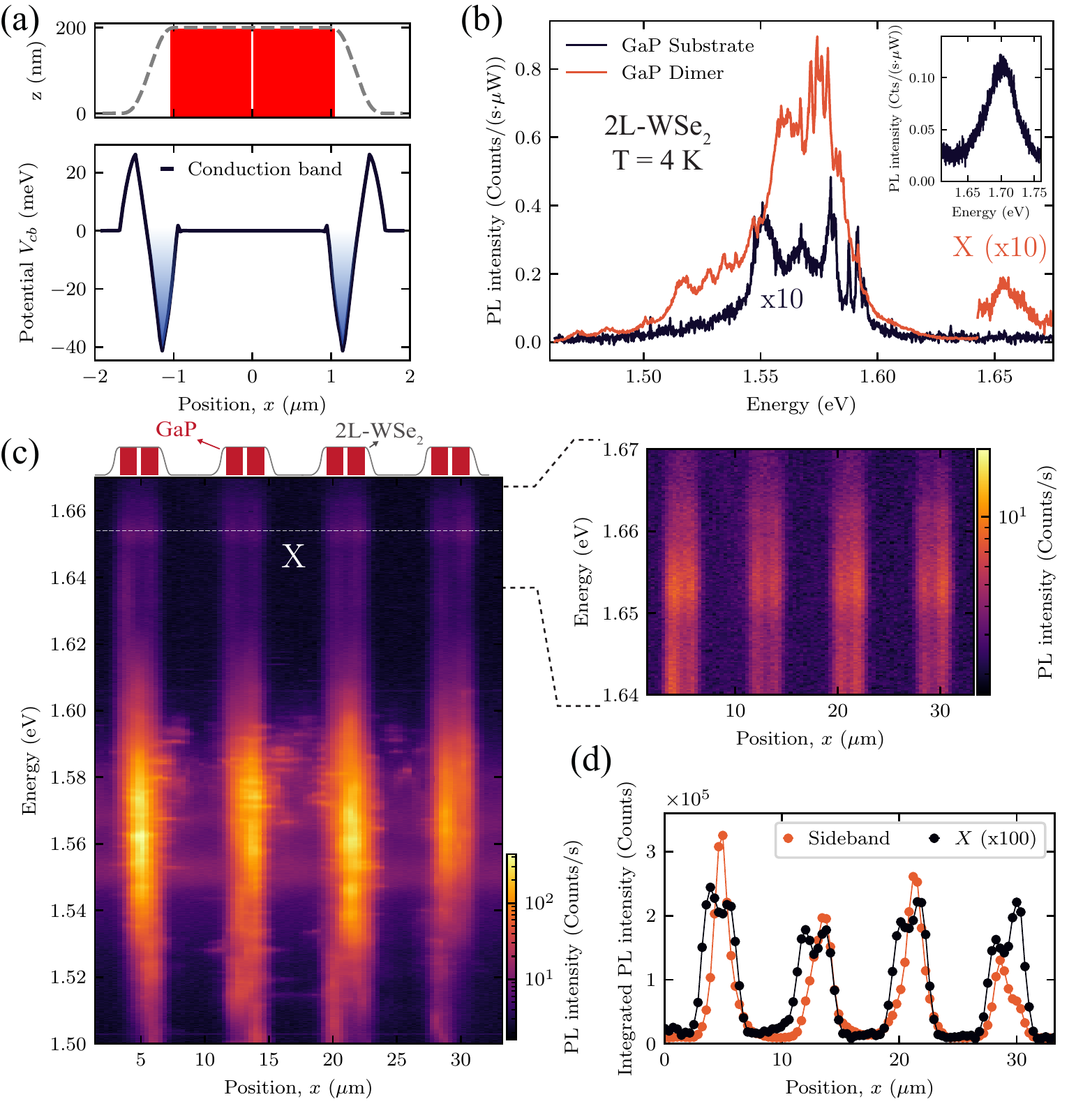}
		\linespread{0.9}
		\caption{
			(a) Top panel: cross-section along the $ x $-axis (as in Fig.1e) of the profile for the 2L-WSe$_{2}$ (dashed line) on top of a dimer nano-antenna (in red, \textit{h} = 200 nm, $ r $ = 500 nm). Bottom panel: strain-induced potential ($ V_{cb} $) calculated for the $K$ valley in the conduction band as a function of $x$. 
			(b)  PL spectra measured at T=4 K using a laser at 685 nm with a power of 100 $\mu$W for 2L-WSe$_{2}$ placed on top of the dimer nano-antenna with $ r=500 $ nm (orange) and on the planar GaP substrate (black). The spectrum measured for the planar GaP is multiplied by 10. For the orange spectrum, the intralayer exciton PL peak ($X$) around 1.65 eV is also multiplied by 10. Inset shows intralayer exciton PL measured for 2L-WSe$_2 $ placed on the planar GaP substrate, exhibiting a maximum around 1.7 eV. This PL spectrum is measured in the micro-PL setup used for room temperature measurements (high collection efficiency), using a 532 nm laser with a high power of 1 mW in order to compensate for an extremely low PL intensity of the intralayer exciton.
			(c) Hyper-spectral image measured at T=4 K using a laser at 685 nm with a power of 100 $\mu$W showing PL for 2L-WSe$_{2} $ deposited on top of four dimer nano-antennas ($ r $ = 500 nm) separated by $ \approx 10$ $ \mu $m (schematically shown at the top of the panel). As shown in the zoomed-in image, the spatial distribution of the intralayer exciton PL shows a clear localization in two spatially separated PL maxima, as predicted by the potential presented in Fig.\ref{fig:fig4}a.
			(d) Integrated PL intensity as a function of \textit{x}. The data in black is for the intralayer exciton with integration limits 1.65-1.66 eV (the PL intensity is multiplied by 100), and in orange for the broad PL band with integration limits 1.56-1.58 eV.}
		\label{fig:fig4}
	\end{figure*}

	From our strain topography model, it is possible to calculate the strain-induced potential for the case of the dimer structures (see Supplementary Note II). We find that the strain-induced potential wells are located at the nano-pillar edges, where the tensile strain is maximized \cite{Feng2012,Brooks2018}. When strain becomes compressive, i.e. where the layer touches the substrate, we observe the presence of potential barriers due to the opposite shift of the band edges, leading to an increase of the bandgap energy.
	Fig.\ref{fig:fig4}a shows the profile of the strained 2L-WSe$_2 $ (dashed line) on top of a dimer nano-antenna with $ r=500 $ nm (red squares). The TMD profile shown in the top panel is calculated for a cross section along the $ x- $axis (as defined in Fig.\ref{fig:fig1}e), where $ x=0 $ is the centre of the dimer gap. We correlate the local deformation with the changes in the energy of the conduction band minimum ($ V_{cb} $), calculated numerically for the K valley (see the bottom panel in Fig.\ref{fig:fig4}a and Supplementary Note II for details).
	The minima of $ V_{cb} $ are localized at the nano-pillar edges, resulting in a confinement potential wells (shaded areas).
	For WSe$ _2 $ the valence band maximum, $ V_{vb} $, exhibits an increase (decrease) in case of tensile (compressive) strain, also resulting in a confinement for holes, although this effect is weaker compared to conduction band electrons. In such a potential landscape, the exciton population will accumulate in the potential wells at the nano-pillar edges.

	Fig.\ref{fig:fig4}b shows a reference PL spectrum of 2L-WSe$_2 $ collected at $T=$4 K on the planar GaP (black) and on a dimer nano-antenna with $ r=500  $ nm (orange). 
	The low temperature PL of 2L-WSe$_2 $ is dominated by the low energy broad band between 1.5 and 1.6 eV, relatively weakly dependent on the strain induced by the dimer. Overall the spectrum is characterized by the non-trivial combination of PL from bright and dark states, phonon replicas \cite{Lindlau2018} and emission of various localized states \cite{Kumar2015a}. The only part of this broad band that consistently shows dependence on the strain induced by the pillars is the lower side of the spectrum between 1.5 and 1.55 eV, which exhibits higher intensity in the strained WSe$_2$. One of the reasons for such behaviour could be that the PL in the range 1.55-1.6 eV may originate from the transitions involving the $Q$ valley in the conduction band \cite{Lindlau2018}, which is weakly sensitive to the strain. On the other hand, we may conclude that the PL in the spectral range of 1.5 to 1.55 eV may arise from the localized exciton states, with the energies following the overall reduction in the direct bandgap by about 45 meV (see details below), or perhaps the optical transitions involving the $K$ valley in the conduction band. However, the complete understanding of this behaviour requires elaborate theory for the optical transitions in bilayer WSe$_2$ and is beyond the scope of this work.
	
	Despite the very low PL intensity, as expected at low $T$ in WSe$_2$, we are still able to observe the momentum-direct $ K\rightarrow K $ exciton transition for the unstrained WSe$_2$ on the planar GaP at 1.7 eV (inset Fig.\ref{fig:fig4}b). As expected for the direct transition, in the strained WSe$_2$ placed on the nano-antenna, the corresponding exciton peak is observed approximately 45 meV lower at 1.655 eV (denoted $ X $). From this energy red-shift we estimate a tensile strain value of $\approx1\%$, consistent with our theoretical model predictions. As we show in Supplementary Note VI, we observe a similar red-shift in the cryogenic PL of the strained 1L-WSe$_2 $ exciton, also in agreement with our model.

	Fig.\ref{fig:fig4}c shows a one-dimensional hyper-spectral PL image of 2L-WSe$_2 $ placed on top of four GaP nano-antennas having $ r=500 $ nm and the gap size varying between 100 and 150 nm from left to right. The PL image is measured by scanning the collection and excitation spot along the nano-antenna $ x $-axis (see Fig.\ref{fig:fig1}e for the axis definition), while on the vertical axis of the figure we report the spectral distribution of the collected PL emission.
	The increased PL intensity corresponds to the enhanced 2L-WSe$_2 $ PL on top of the dimer nano-antennas, with a negligible signal collected on the flat GaP (dark areas in between). 
	
	In the right panel of Fig.\ref{fig:fig4}c we zoom in on the direct exciton PL emission in the strained WSe$_2$ at around 1.655 eV. The exciton PL intensity exhibits a clear localization into two spatially separated maxima. Fig.\ref{fig:fig4}d shows the integrated intensity of the exciton PL (in black) clearly highlighting this behaviour. Such two-peaked distribution is also reproduced to some degree for the lowest part of the PL spectrum below $\approx$ 1.52 eV. This is a consequence of the sensitivity of this part of the spectrum to the induced strain as observed in Fig.\ref{fig:fig4}b. We conclude that for the two types of states, exciton relaxation into the strain-induced potential minima is significant.
	
	As seen both in Fig.\ref{fig:fig4}c and d, such two-peaked distribution is not replicated clearly for the broad band between 1.55 and 1.6 eV, which is consistent with a weak sensitivity of the involved electronic states to the strain observed in Fig.\ref{fig:fig4}b.

	\section*{Discussion}
	
	In summary, we have demonstrated strain-tuning of the electronic band structure in 2D semiconductor WSe$_2$ by placing the atomic mono- and bilayers on pre-patterned dielectric nano-antennas. 
	Because of the efficient coupling with the optical mode confined at the surface of the nano-antenna, the PL of the strained WSe$_2$ is enhanced. It can thus be clearly detected with negligible contribution from the PL of the unstrained material, that would normally mask the effect of strain, for example, in the case of the pillars made of SiO$_2$ or polymers, which were employed previously to induce strain \cite{Palacios-Berraquero2016c,Li2015}.
	We therefore can clearly deduce the amount of induced strain from the red-shift in WSe$_2$ PL spectra at room temperature in agreement with our theory. Moreover, we observe excitons confined into strain-induced potential wells at cryogenic temperature, the presence of which is also predicted by our theoretical model. 
	
	The results presented in this work open the way for the use of strain as an additional degree of freedom in engineering of the light-matter interaction in 2D materials, and could find applications for studying 2D excitons in confined potentials, relevant for positioning of strain-induced WSe$_2 $ quantum emitters\cite{Palacios-Berraquero2016c,Branny2016}, as well as for Bose-Einstein condensation of interlayer excitons in TMDs heterostructures\cite{Wang2019b}.
	
	Additionally, our continuum-mechanical theoretical approach can be expanded to describe the strain distribution and resulting distortion potential for 2D TMDs coupled to a broad range of nano-structures including plasmonic nano-antennas. This opens the way for designing optically active nano-photonic platforms interfaced with strained 2D materials and their heterostructures, with prospects in photodetection, light emission and photovoltaic applications. 
	
	\bigskip
	\noindent
	\textbf{Methods}
	
	\noindent
	\textbf{Sample fabrication}
	\noindent
	The GaP nano-antennas are fabricated with a top-down lithography process, as described in Ref.\cite{Cambiasso2017a}. The monolayers and bilayers of WSe$ _{2} $ were mechanically exfoliated from commercially available bulk crystal (HQGraphene) and the layer thickness identified via a PL imaging technique\cite{Alexeev2017}. The 2D layers are transferred on top of the GaP nano-antennas in a home-built transfer setup, with an all-dry transfer technique\cite{Castellanos-Gomez2014a}.
	
	\noindent
	\textbf{Optical spectroscopy}
	\noindent
	The PL image used in Fig.1d is obtained in a commercial bright-field microscope (LV150N Nikon), with the technique described in Ref.\cite{Alexeev2017},  where the white light source is used both to image the TMD layers and to excite its PL emission. A 550 nm short-pass filter is placed in the illumination beam path to remove the low energy side of the white light emission spectrum. A high numerical aperture objective (Nikon x100 NA=0.9) is used to direct the light on the sample and for collecting the reflected and emitted light. In the collection path a 600 nm long-pass filter rejects the white light source, while the PL image from the TMD layer is detected with a microscope camera (DS-Vi1, Nikon).
	
	Room temperature optical spectroscopy is performed in a home built micro-PL setup in  back reflection geometry. A diode laser with a wavelength of $\lambda_{exc}=685$nm is used as the excitation source and directed into an infinity corrected objective (Mitutoyo 100x NA=0.7), allowing focussing in a 3 $\mu$m spot.  We use a laser power of 30 $\mu$W. The sample is placed on a motorized stage (STANDA-8MTF) which allows the automated mapping of the sample surface. 
	The emitted light is collected by the same objective and using free space (no optical fibers) is then coupled to a spectrometer  (Princeton Instruments SP2750) equipped with a high-sensitivity liquid nitrogen cooled charge-coupled device (Princeton Instruments PyLoN).
	
	For optical spectroscopy at cryogenic temperature T $\approx$4 K, the sample is placed into a microscope stick (attocube systems AG) with a cage structure and a window for optical access. The sample is placed onto piezoelectric nano-positioners (attocube ANP101 models). The cage structure is fitted into an aluminium tube and held under high-vacuum. The tube is then inserted in a liquid helium bath cryostat and an optical breadboard with a micro-PL setup is placed on top of the tube to carry out the optical analysis. PL spectra were excited with the same laser as for room temperature measurements with $\lambda_{exc}=685$nm. The incident power was kept at 100 $\mu$W, and the excitation spot with a diameter of 3 $\mu$m was achieved. The light emitted by the sample is coupled to a single mode optical fibre and detected in the same spectrometer/CCD system employed for the room temperature spectroscopy.
	
	Additional measurements at a cryogenic temperature of T$\approx$10 K reported in the inset of Fig.4b were carried out in a He flow cryostat using the same setup as was employed for the room temperature PL measurements. The PL was excited with a 532 nm laser with 1 mW power and coupled via free space into the spectrometer.
	
	\begin{acknowledgement}
		L. S. and A. I. T. thank the financial support of the European Graphene Flagship Project under grant agreements 785219, and EPSRC grant EP/S030751/1. L. S., A. I. T., M. B. and G. B. thank the European Union's Horizon 2020 research and innovation programme under 
		ITN Spin-NANO Marie Sklodowska-Curie grant agreement no. 676108. P. G. Z. and A. I. T. thank the European Union's Horizon 2020 research and innovation programme under ITN 4PHOTON Marie Sklodowska-Curie grant agreement no. 721394. A. G. and A. I. T. acknowledge funding by EPSRC grant EP/P026850/1. J. C., S. M., S. A. M., and R. S. acknowledge funding by EPSRC (EP/P033369 and EP/M013812). S. A. M. acknowledges the Lee-Lucas Chair in Physics and the Solar Energies go Hybrid (SolTech) programme.
		
		\noindent
		\textbf{Author contributions}
		
		\noindent
		L. S., A. I. T., S. A. M., R. S. conceived the idea of the experiment. L. S. and P. G. Z. fabricated WSe$_2$ layers and transferred them on the GaP nano-antennas. L. S., P. G. Z. and A. G. carried out microscopy and optical spectroscopy measurements on WSe$_2$. J. C. fabricated GaP nano-antennas. S. M., J. C. and R. S. designed GaP nano-antennas. L. S., P. G. Z., A. G. and A. I. T. analysed optical spectroscopy data. M. B. and G. B. developed the theory and carried out simulations for strained WSe$_2$. G. B., S. A. M., R. S. and A. I. T. managed various aspects of the project. L. S., M. B. and A. I. T. wrote the manuscript with contributions from all co-authors.  A. I. T. oversaw the whole project.
	\end{acknowledgement}
	
	\begin{suppinfo}
		
		\begin{itemize}
			\item Optical resonances of GaP dimer nano-antennas; theoretical modelling of strain topography; Raman spectroscopy of strained 1L-WSe$ _2 $, Raw PL spectra from strained WSe$ _2 $; PL spectra for strained 2L-WSe$ _2 $; Strain-induced exciton red-shift in 1L-WSe$ _2 $ at cryogenic temperatures;
		\end{itemize}
		
	\end{suppinfo}
	
	
	\providecommand{\latin}[1]{#1}
	\makeatletter
	\providecommand{\doi}
	{\begingroup\let\do\@makeother\dospecials
		\catcode`\{=1 \catcode`\}=2 \doi@aux}
	\providecommand{\doi@aux}[1]{\endgroup\texttt{#1}}
	\makeatother
	\providecommand*\mcitethebibliography{\thebibliography}
	\csname @ifundefined\endcsname{endmcitethebibliography}
	{\let\endmcitethebibliography\endthebibliography}{}

\pagebreak
\renewcommand{\figurename}{SUPPLEMENTARY FIGURE}

\setcounter{figure}{0}

	\section*{Supplementary note I: optical resonances of GaP dimer nano-antennas}
	
	\begin{figure}[h]
		\centering
		\includegraphics[width=0.7\linewidth]{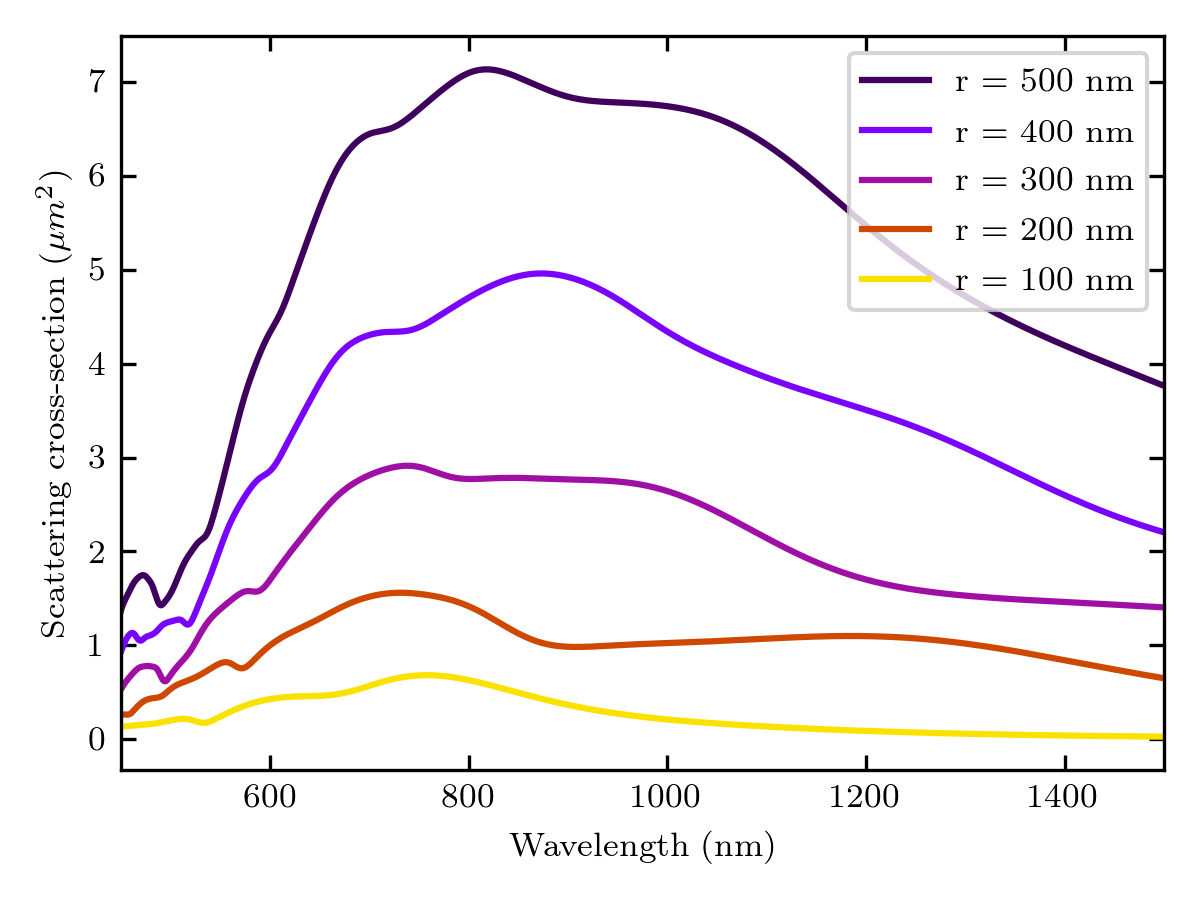}
		\caption{Simulated scattering cross section of GaP dimer nano-antennas with different radii (height = 200 nm, gap width = 50 nm), excited with a plane wave linearly polarized along the x-axis as defined in Fig.1 of the main text.}
		\label{fig:figsscattering}
	\end{figure}
	
	\newpage
	\pagebreak

	\section*{Supplementary note II: theoretical modelling of strain topography}
	
	The strain topography of stretched transition metal dichalcogenides (TMD) layers is achieved by modelling the out-of-plane displacement relative to the underlying substrate. We applied a continuum-mechanical plate-theory approach, as introduced in Ref.\cite{Brooks2018}, with the following reduced form of the F\"{o}ppl-von K\'{a}rm\'{a}n equations, valid for such passively strained low-dimensional systems \cite{Landau}:
	
	\begin{equation}\label{press}
		D\Delta^{2}\zeta- P = 0
	\end{equation}
	
	\bigskip
	
	\noindent where $ P $ is the pressure the atomic layer experiences, $ \zeta $ is the function that describes the height profile of the layer \cite{Brooks2018}, $ \Delta^{2} $ is the biharmonic operator, given by the square of the Laplacian $ \Delta $, and $ D $ is the flexural rigidity of the TMD layer defined as:
	
	\begin{equation}\label{rigidity}
		D= \dfrac{Eh^{3}}{12(1 - \sigma^{2})}
	\end{equation}
	
	\bigskip
	
	\noindent here $ E $ is the Young's modulus, $ \sigma $ is the Poisson’s ratio and $ h$ is the TMD layer thickness.

	To model the WSe$ _2$ strain topography above the dimer structure, we found a solution to Eq.\ref{press}. The model of pressure \cite{Brooks2018} adopted here, consistent with the microscopy measurements, is the plate model (P = 0) with boundary conditions:
	
	\begin{eqnarray}
		\zeta_{P=0}(r) &= H \\
		\zeta_{P=0}(r_T) &= 0 \\
		\partial_{\rho}\zeta_{P=0}|_{r,r_T} &= 0
	\end{eqnarray}
	
	\bigskip
	\begin{figure}[b!]
		\centering
		\includegraphics[width=0.7\linewidth]{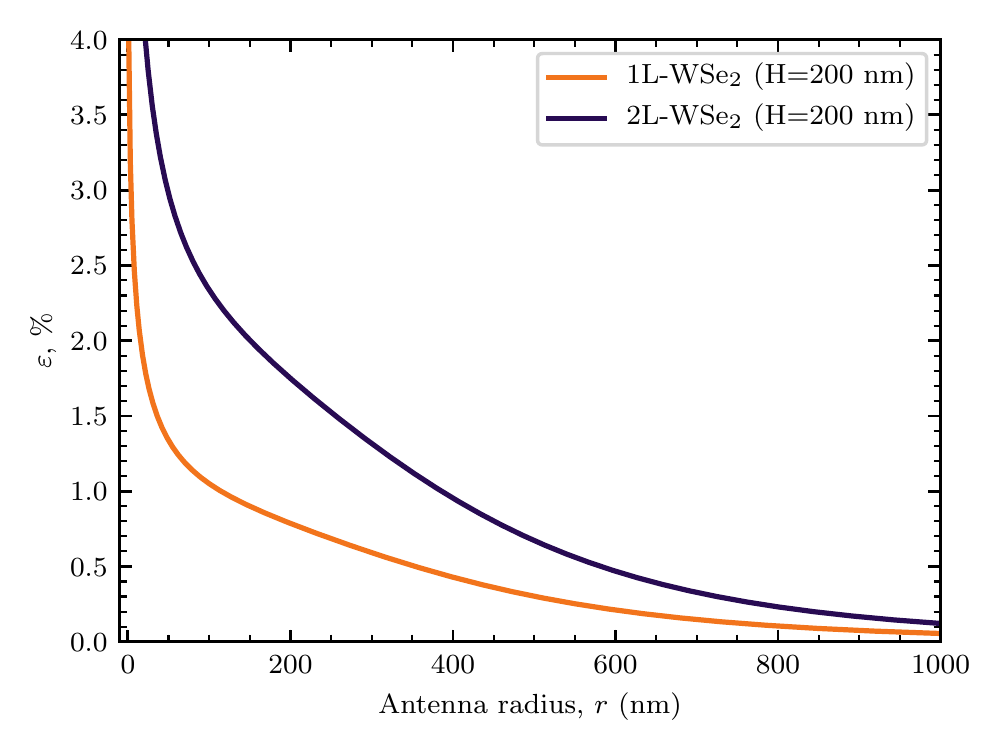}
		\caption{Uniaxial tensile strain values as a function of the nano-pillar radius, calculated at the topmost edge of the dimer nano-antenna (H = 200  nm) for both single and double layer WSe$_2 $.}
		\label{fig:2c}
	\end{figure}
	\noindent where $ H$ is the height of the dimer antenna, $ r $ is the nano-pillar radius and $ r_T $ is the tenting radius, the distance between the antenna pillars and the point where the WSe$ _2$ layer meets the substrate. 
	We obtain the following analytical solution for the height field, relative to a single pillar, in rotational symmetry:
	
	\begin{align}
		\zeta_{P=0}(\rho)=\frac{H}{\left(r_T^2-r^2\right){}^2-4 r_T^2 r^2 \ln ^2\left[\frac{r_T}{r}\right]}  \left\{2 r_T^2 \ln [r_T] \left(r^2+2 r^2 \ln \left[\frac{r}{r_T}\right]-\rho^2\right)\right.&\\\left.+\left(\rho^2-r_T^2\right) \left(r^2+2 r^2 \ln \left[r\right]-r_T^2\right)\right.&\\\left.+2 \ln [\rho] \left(\rho^2 \left(r_T^2-r^2\right)+2 r_T^2 r^2 \ln \left[\frac{r_T}{r}\right]\right)\right\}
		\label{eq:0PZeta}
	\end{align}
	
	\bigskip
	
	Taking this approximation as a starting point, we obtain a numerically averaged height field over the dimer geometry by taking values of $ r_T $ from microscopy measurements, and interpolation in the gap region where the rotational symmetry of the approximate model is violated. From a satisfactory model of the height field, the strain component responsible for the band gap renormalization is given by the trace of the strain tensor $ \varepsilon_{ij} $, given as\cite{Pearce2016}:
	
	\begin{equation}
		\mathcal{T}=Tr[\varepsilon_{ij}]=\dfrac{(2\sigma-1)h}{1-\sigma}\Delta\zeta
	\end{equation}
	
	\bigskip
	
	We resolved a relationship between the tensile strain maximum and the nano-pillar radius. Fig.S\ref{fig:2c} shows the calculated strain values at the topmost edge of a nano-antenna ($ H=200 $ nm), as a function of the nano-pillar radius. For both 1L and 2L-WSe$ _2 $ we observe an increased strain value as \textit{r} decreases, with a stronger magnitude for the bilayer. This trend can be explained as the assumed model approaches a deflection of a Dirac-delta like deformation, when $ r\rightarrow 0 $, giving an undefined pole in the strain value. An increased strain in bilayers is directly related to its larger rigidity (Eq.\ref{rigidity}).
	
	\begin{figure}[t!]
		\centering
		\includegraphics[width=0.7\linewidth]{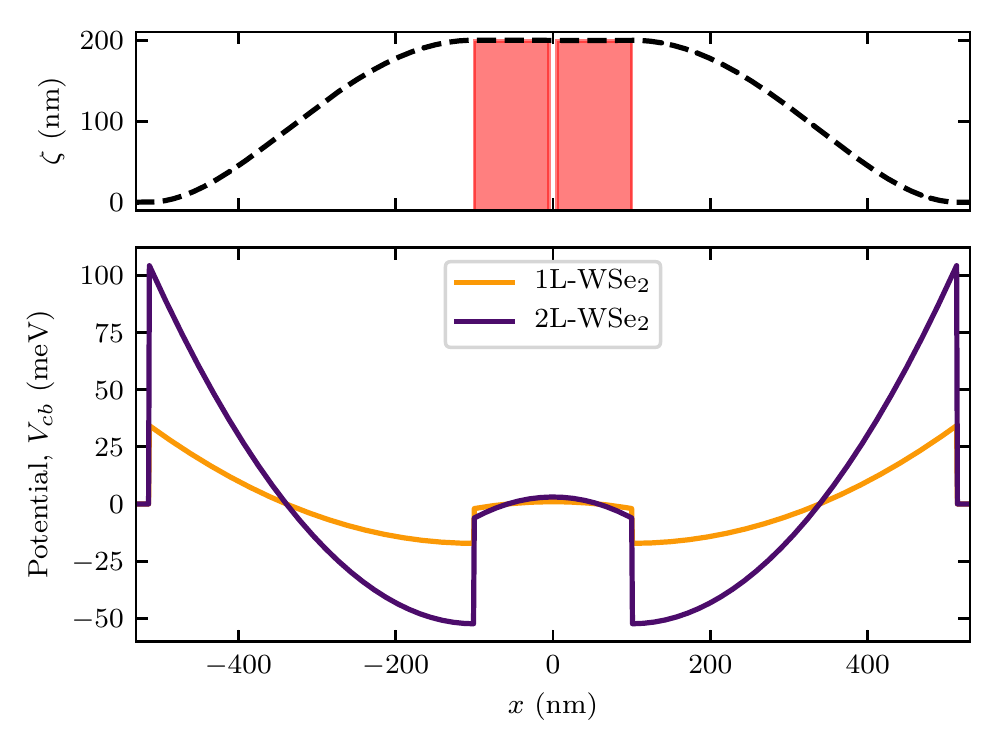}
		\caption{Top panel: Height field profile (black dashed line) for a dimer with $ H = 200 $ nm and $ r = 50 $ nm, schematically shown in red. Bottom panel: conduction band potential minimum ($ V_{cb} $), calculated for the K(K') point of the Brillouin zone, for both single (1L) and double (2L) layer WSe$ _2 $, relative to the height field profile shown in the top panel. The compressive (tensile) strain region is correlated to an increase (decrease) of the potential value.}
		\label{fig:5a}
	\end{figure}

	After defining a strain topography, it is possible to calculate a strain-induced deformation potential in the WSe$ _2 $ band structure, following a tight-binding approach\cite{Pearce2016,Brooks2018}, as:
	
	\begin{equation}
		V = 
		\begin{pmatrix}
			V_{vb}  &  0 \\
			0 & V_{cb}
		\end{pmatrix}
		=
		\begin{pmatrix}
			\delta_{v}\mathcal{T}  &  0 \\
			0 & \delta_{c}\mathcal{T}
		\end{pmatrix}
	\end{equation}
	
	\bigskip
	
	\noindent where $\delta_{c}$ and $\delta_{v}$ are the parameters governing the strain response for the conduction ($ V_{cb} $) and valence band ($ V_{vb} $) at the K(K') point, respectively. 
	Fig.S\ref{fig:5a} shows the calculated conduction band deformation potential, for both 1L and 2L-WSe$ _{2} $, on a dimer structure with $ r=50 $ nm and $ H=200 $ nm, along the line connecting the centre of the two nano-pillars. We found a stronger modulation for the bilayer, as expected from the higher strain values shown in Fig.S\ref{fig:2c}.
	The potential profile is correlated with the local reduction (increase) of WSe$ _{2} $ bandgap under tensile (compressive) strain. As such, the maximum tensile deformation, located at the nano-antenna edges, corresponds to a decrease in the conduction band minimum. In correspondence with the compressive strain area, where the layers meet the substrate, the bandagap energy is increased resulting in a potential barrier for free excitons.

	\pagebreak
	\section*{Supplementary note III: Raman spectroscopy of strained 1L-WSe$ _2 $}
	
\begin{figure}[h!]
		\centering
		\includegraphics[width=0.5\linewidth]{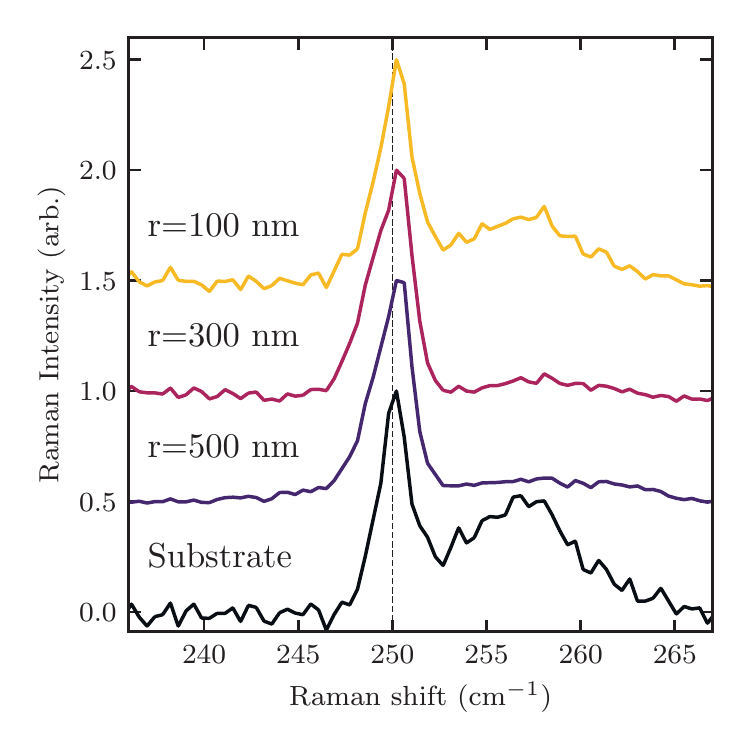}
		\caption{Raman scattering spectra for 1L-WSe$ _2 $ placed on the planar GaP substrate and on different radii nano-antennas. The presence of strain is associated with a shift to higher wavenumber of the main peak at 250 cm$ ^{-1} $, corresponding to the $ E' + A'_1 $ mode. The spectra are normalized and shifted vertically for display.}
		\label{fig:figs3raman}
	\end{figure}

	Figure S\ref{fig:figs3raman} shows the Raman scattering signal from a monolayer of WSe$ _2 $ transferred on top of GaP nano-antennas with different radii, compared to the signal from the planar GaP substrate (black). The 1L-WSe$ _2 $ peak at 250 cm$ ^{-1} $ exhibits a strain-induced shift when deposited on the GaP nano-antennas. The results are consistent with those reported in Ref.\cite{Desai2014}.  The Raman scattering signal is collected in the setup described in Ref.\cite{Sortino2019}. 
	
	\newpage
	\pagebreak
	
	\section*{Supplementary note IV: Raw PL spectra from strained WSe$ _2 $}
	
	\begin{figure}[h]
		\centering
		\includegraphics[width=1\linewidth]{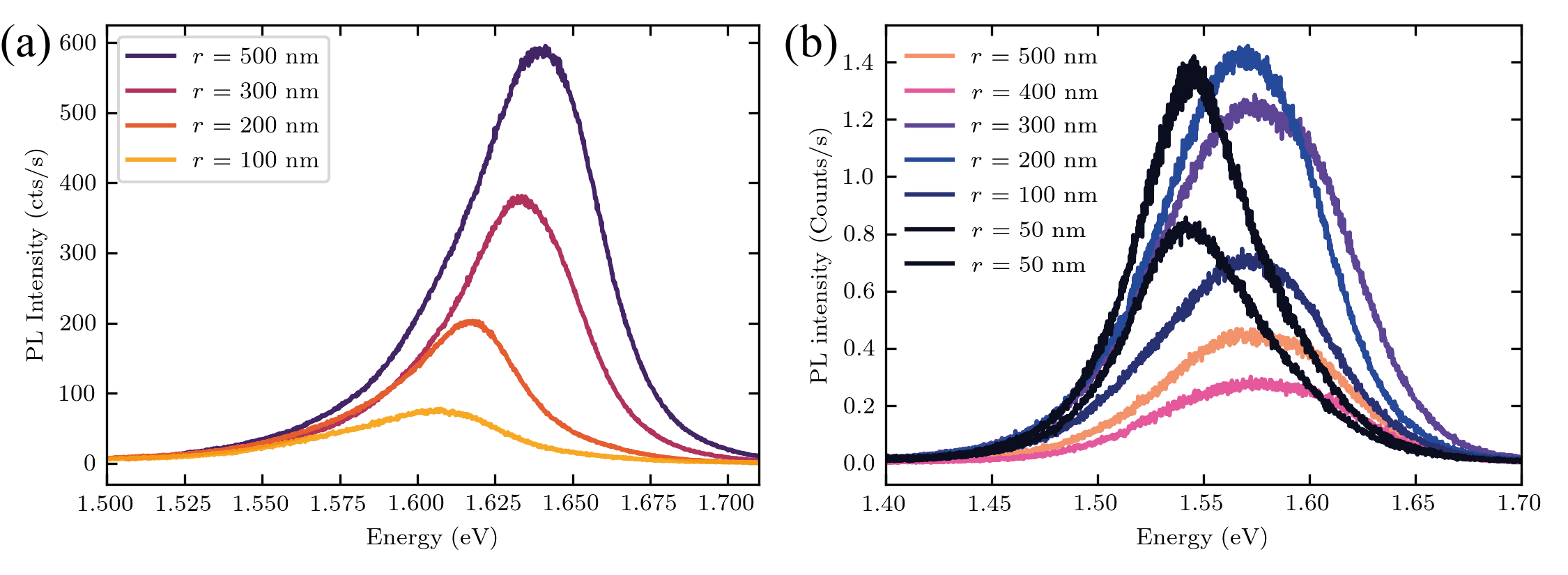}
		\caption{(a) Raw PL spectra for 1L-WSe$ _2 $ placed on GaP dimer nano-antennas with different radii. (b) Raw PL spectra for 2L-WSe$ _2 $ placed on GaP dimer nano-antennas with different radii. The spectra are collected under the same experimental condition.}
		\label{fig:figsplnonorm}
	\end{figure}
	
	\newpage
	\pagebreak
	
	\section*{Supplementary note V: PL spectra for strained 2L-WSe$ _2 $}
	
\begin{figure}[b!]
		\centering
		\includegraphics[width=1\linewidth]{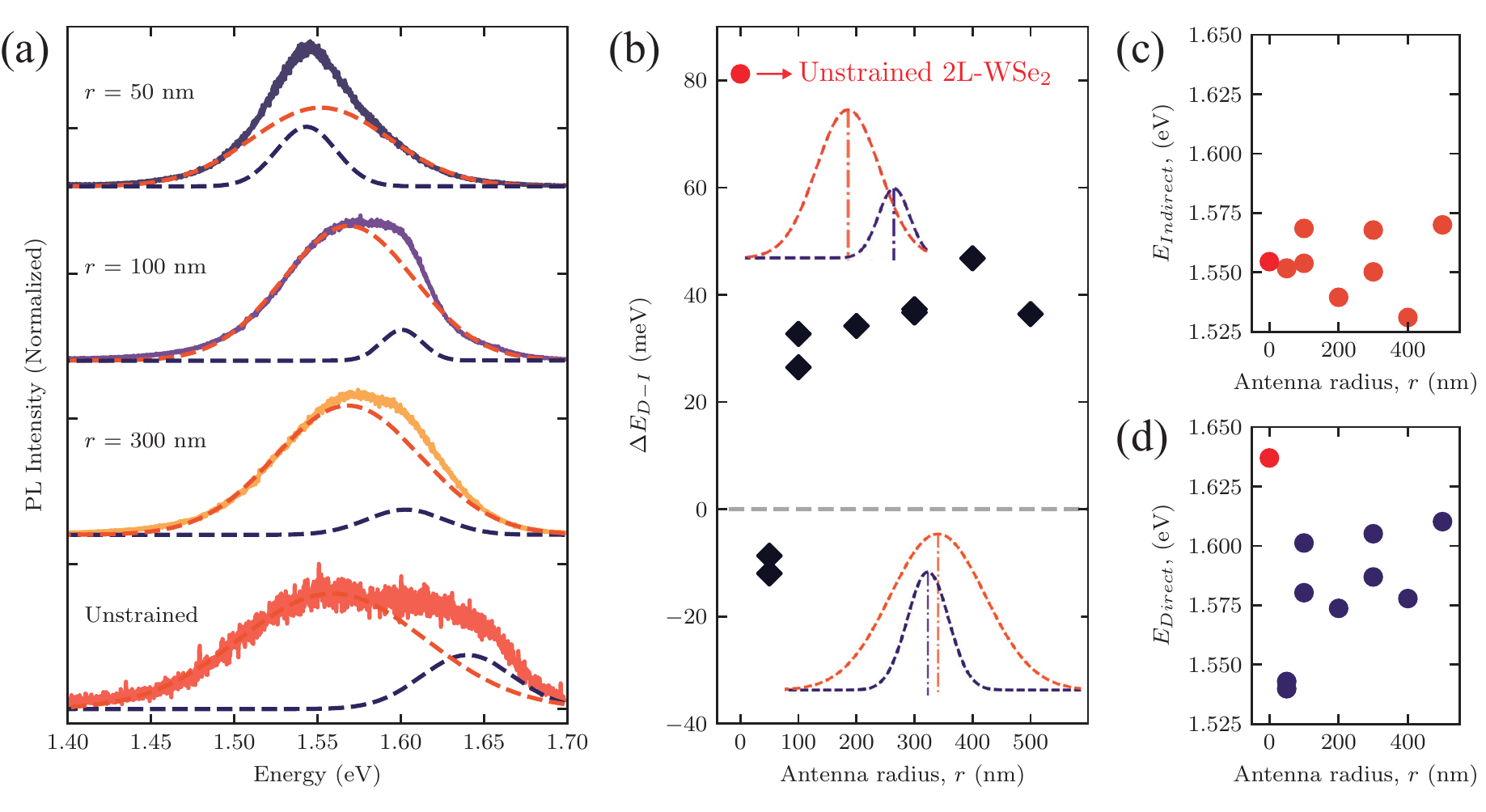}
		\caption{(a) 2L-WSe$ _2 $ PL emission collected on nano-antenna with different radius. The spectra are fitted with two Gaussian peaks relative to the direct (purple) and indirect (orange) transitions. (b) Energy difference ($ \Delta E_{D-I}$) between the direct and indirect peak extracted from the PL emission of 2L-WSe$ _2 $ on nanoantennas with different radii. In red the value extracted from unstrained 2L-WSe$ _2 $ on planar GaP. The dashed line marks the transition to direct bandgap. (c,d)  Direct and indirect peak position (in red for unstrained 2L-WSe$ _2 $). No evident change is found in the $ E_{Indirect} $ peak (c).  In contrast, the $ E_{Direct} $ position (d) shows a clear modulation. This effect confirms that a transition to a direct bandgap is caused by the reduction in energy of the $ K\rightarrow K $ recombination pathway.}
		\label{fig:2l-fit-qk}
	\end{figure}

	To discriminate the relative role of the competing valleys in the transition to a direct bandgap in strained 2L-WSe$ _2 $, we extracted the relative emission peak maxima from the PL spectra (as described in Fig.3 of the main text) when deposited on nano-antennas with different radii (Fig.S\ref{fig:2l-fit-qk}a). The spectra are fitted with two Gaussian functions from which we have extracted their relative energy difference, $ \Delta E_{D-I} = E_{Direct} - E_{Indirect}$ (Fig.S\ref{fig:2l-fit-qk}b). 
	For the unstrained bilayer on the planar GaP substrate we found the largest value of $\Delta E_{D-I}\approx80 $ meV. When deposited on the nano-antennas, $ \Delta E_{D-I}$ exhibits an initial reduction from the unstrained values, further exhibiting a shift of $ >30 $ meV when approaching the nano-antenna with $ r=50 $ nm.
	Figures S\ref{fig:2l-fit-qk}c-d show the position of each peak maxima position, as extracted from the 2L-WSe$ _2 $ PL spectra collected on nano-antennas with different radii. As the nano-pillar radius is reduced, thus increasing the strain magnitude as shown in Fig.S\ref{fig:2c}, the $ E_{Indirect} $ peak does not exhibits a clear shift in its position when compared to the unstrained case (red circle). On the other hand, we found a large shift of the $ E_{Direct} $, up to a value of approximately 100 meV from the unstrained case. This behaviour confirms the predominant role of the $ E_{Direct} $ peak in the indirect-to-direct bandgap transition in 2L-WSe$ _2 $ under strain.

	\pagebreak

	\section*{Supplementary note VI:  Strain-induced exciton red-shift in 1L-WSe$ _2 $ at cryogenic temperatures}

	\begin{figure}[h!]
		\centering
		\includegraphics[width=1\linewidth]{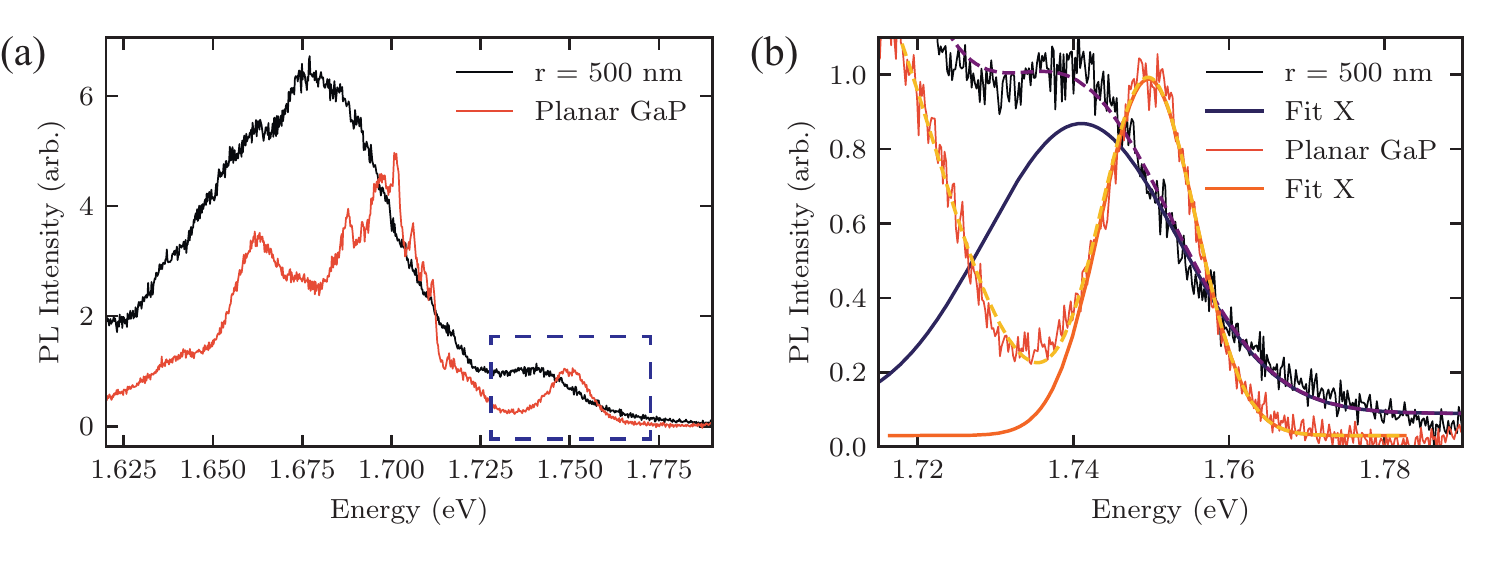}
		\caption{(a) PL emission of 1L-WSe$ _2 $, collected at T=4 K, on the planar GaP substrate (orange) and on a $ r=500  $ nm dimer nano-antenna (black). The spectra are normalized to the intensity of the direct exciton emission peak (dashed line box). (b) Gaussian fit of the direct exciton peak for both spectra shown in Fig.S\ref{fig:SI_exciton}a (cumulative fit shown with the dashed line). A red-shift of $ \approx 10 $ meV is observed when on top of the GaP nano-antenna with $ r=500 $ nm, consistent with the values extracted at room temperature and with the prediction of our theoretical model.
		}
		\label{fig:SI_exciton}
	\end{figure}

\end{document}